\documentclass{article} 
\newcommand{\beq}{\begin{eqnarray}}
\newcommand{\eeq}{\end{eqnarray}}
\newcommand{\beqnn}{\begin{eqnarray*}}
\newcommand{\eeqnn}{\end{eqnarray*}}
\newcommand{\rd}{\partial}
\newcommand{\tp}[1]{\,{}^{\mathrm{t}}#1}

\newcommand{\PP}{\mathbf{P}}

\newcommand{\ee}{\mathbf{e}}
\newcommand{\xx}{\mathbf{x}}
\newcommand{\yy}{\mathbf{y}}
\newcommand{\calL}{\mathcal{L}}
\newcommand{\SU}{\mathrm{SU}}
\begin{document}

\title{Integrable systems whose spectral curve
is the graph of a function}
\author{Kanehisa Takasaki\\
\normalsize Department of Fundamental Sciences, \\
\normalsize Faculty of Integrate Human Studies, Kyoto University\\
\normalsize Yoshida, Sakyo-ku, Kyoto 606-8501, Japan\\
\normalsize E-mail: takasaki@math.h.kyoto-u.ac.jp}
\date{}
\maketitle

\begin{abstract}
For some integrable systems, such as the open Toda 
molecule, the spectral curve of the Lax representation 
becomes the graph $C = \{(\lambda,z) \mid z = A(\lambda)\}$ 
of a function $A(\lambda)$.  Those integrable systems provide 
an interesting ``toy model'' of separation of variables.  
Examples of this type of integrable systems are presented 
along with generalizations for which $A(\lambda)$ lives 
on a cylinder, a torus or a Riemann surface of higher genus.  
\end{abstract}

\begin{flushleft}
\end{flushleft}

\begin{flushleft}
nlin.SI/0211021
\end{flushleft}


\section{Introduction}

Separation of variables (SOV) is an old and new subject, 
which was born in the Hamilton-Jacobi theory of 
the nineteenth century \cite{bib:St}, 
revived by Moser in the light of the modern theory 
of integrable systems \cite{bib:Mo-SOV}, and extended 
by Sklyanin and his collaborators to a wide range of 
classical and quantum integrable systems \cite{bib:Sk}.  
Moreover, recent studies have revealed a number of 
new aspects of SOV in the context of 
bi-Hamiltonian structures \cite{bib:Mo-To,bib:Bl,bib:Ma-etal}, 
universal Hamiltonian structures \cite{bib:Kr-Ph}, 
algebraic surfaces \cite{bib:Hu,bib:Bo}, 
geometric Langlands duality \cite{bib:Fr}, 
string theory \cite{bib:Go-Ne-Ru}, 
affine Jacobi varieties \cite{bib:Na-Sm,bib:Sm-Ze}, etc.  

This article is a supplement to the joint work with 
Takebe \cite{bib:Ta-Ta} on SOV of integrable systems. 
This work was motivated by a rather naive question --- 
what will be the simplest nontrivial examples of 
SOV of integrable systems?   The first plan of this 
research was to use such systems as a ``toy model'' 
for better understanding of Sklyanin's work on 
SOV of quantum integrable systems (in particular, 
Calogero-Moser and Ruijsenaars-Schneider systems 
\cite{bib:Sk-etal}).  Unfortunately, 
the program on quantum integrable systems has been 
unsuccessful due to technical difficulties.  
However, as we shall see below, searching for simplest 
examples of SOV is an interesting issue in itself.

\section{Separation of variables of integrable Hamiltonian systems}

Let us consider a Hamiltonian system with canonical 
variables $q_1,\ldots,q_n,p_1,\ldots,p_n$ and 
a Hamiltonian $H = H(q_1,\ldots,q_n,p_1,\ldots,p_n)$. 
This system is said to be integrable 
(or completely integrable) if it has 
$n$ functionally independent (i.e., 
$dH_1\wedge \cdots \wedge dH_n \not= 0$) 
first integrals $H_1,\ldots,H_n$ in involution 
(i.e., $\{H_j, H_k\} = 0$).  The joint level set 
of the first integrals, 
\beq
  H_\ell(q_1,\ldots,q_n,p_1,\ldots,p_n) = E_\ell, 
  \quad \ell = 1,\ldots,n, 
\eeq
is a Lagrangian submanifold of the $2n$-dimensional 
phase space, which (after canonical transformations 
of coordinates, if necessary) can be expressed as 
\beq
  p_j = \rd S/\rd q_j 
\eeq
by a function $S = S(q_1,\ldots,q_n,E_1,\ldots,E_n)$.  
In the traditional terminology of the Hamilton-Jacobi theory, 
$S$ is a ``complete solution'' of the joint 
Hamilton-Jacobi equations 
\beq
  H_\ell(q_1,\ldots,q_n,\rd S/\rd q_1,\ldots,\rd S/\rd q_n) 
  = E_\ell, 
  \quad \ell = 1,\ldots,n. 
\eeq
The system can be thereby transferred to an action-angle 
systems with the action variables $E_\ell$ and  the angle 
variables $\phi_\ell = \rd S/\rd E_\ell$.  

In this setup, SOV means that the equations of 
the joint level set takes the ``separated'' form 
\beq
  f_j(\lambda_j,\mu_j, E_1,\ldots,E_n) = 0, 
  \quad j = 1,\ldots,n, 
  \label{eq:sep-level-set}
\eeq
in another set of Darboux coordinates 
$\lambda_1,\ldots,\lambda_n,\mu_1,\ldots,\mu_n$ 
(``separation variables''). 
One can then assume the generating function $S$ 
in the separated form 
\beq
  S = \sum_{j=1}^n S_j(\lambda_j, E_1,\ldots,E_n) 
\eeq
and obtain each term $S_j$ by solving the ODE 
\beq
  f_j(\lambda_j,dS_j/d\lambda_j, E_1,\ldots,E_n) = 0. 
\eeq

Note that the existence of first integrals itself 
tells nothing about separability; a prescription 
of SOV has to be sought elsewhere.  Sklyanin's 
``magic recipe'' \cite{bib:Sk} is a prescription 
based on the Lax representation of integrable systems.  
The idea stems from the work of the Montreal group 
on generalized Moser systems \cite{bib:AHH}, 
and is also closely related to the algebro-geometric 
method of the Moscow group \cite{bib:Du-Kr-No}. 
Suppose that that the integrable system has 
a Lax representation of the form 
\beq
  \dot{L}(z) = [L(z), M(z)]
\eeq
with $N \times N$ matrices $L(z)$ and $M(z)$ 
that depends on a spectral parameter $z$.  
The characteristic polynomial 
\beq
  f(\lambda,z) = \det(\lambda I - L(z)) 
  = \lambda^N + \sum_{\ell=1}^N p_\ell(z)\lambda^{N-\ell} 
\eeq
of $L(z)$ does not depend on the time variable. 
A set of first integrals in involution can be 
obtained from the coefficients of expansion 
of $p_\ell(z)$'s by a suitable basis of functions 
of $z$.  The curve 
\beq
  C = \{(\lambda,z) \mid f(\lambda,z) =  0\} 
\eeq
defined by $f(\lambda,z)$ is called the 
``spectral curve'' of the $L$-matrix $L(z)$.  
At each point $(\lambda,z)$ of $C$ is sitting 
the eigenspace of $L(z)$ with eigenvalue $\lambda$. 
For many integrable systems, this eigenspace is 
one-dimensional at almost all points of $C$, 
and the components of a suitably normalized eigenvector 
$\phi(\lambda,z)$ are meromorphic functions on $C$ 
with a finite number of poles.  It is the coordinates 
$\lambda_j,z_j$ of these poles that play the role 
of separation variables in Sklyanin's magic recipe. 
These variables are constrained by the equation 
\beq
  f(\lambda_j,z_j) = 0 
\eeq
of the spectral curve, and these equations 
can be interpreted as the separated equations 
(\ref{eq:sep-level-set}) of the joint level set 
of first integrals. Moreover, in most cases, 
the Poisson brackets of these separation 
variables take the separated form 
\beq
  \{\lambda_j,\lambda_k\} = \{z_j,z_k\} = 0, 
  \quad 
  \{\lambda_j,z_k\} = \delta_{jk}g(\lambda_k,z_k), 
\eeq
where $g(\lambda_k,z_k)$ is a common function 
of $\lambda_k$ and $z_k$ (typically, 
$1, \lambda_k, z_k$ or $\lambda_k z_k$).  
In other words, the symplectic form of the system 
can be written as 
\beq
  \Omega 
  = \sum_{j=1}^N \frac{dz_j \wedge d\lambda_j}{g(\lambda_j,z_j)}. 
\eeq

\section{If the spectral curve is the graph of a function} 

Our concern lies in the case where the spectral curve 
becomes the graph of a function $A(\lambda)$:  
\beq
  C = \{(\lambda,z)\mid z = A(\lambda)\}. 
\eeq
Here is a list of examples of integrable systems of this type: 
\begin{enumerate}
\item Open Toda molecule
\item Open Ruijsenaars-Toda molecule 
\item Rational and trigonometric Calogero-Moser systems 
\item Rational and trigonometric Ruijsenaars-Schneider systems 
\end{enumerate}
These integrable systems are usually understood 
to be associated with no spectral curve, 
because the usual Lax representation does not 
contain a spectral parameter.  The fact is that 
they do have a Lax representation with a spectral 
parameter, and that the spectral curve takes 
the form mentioned above.  

SOV of these examples is almost parallel to 
more standard cases.  The function $A(\lambda)$ 
is expanded to a linear combination of a suitable basis 
$f_0(\lambda),f_1(\lambda),\ldots,f_N(\lambda)$ 
of functions as 
\beq
  A(\lambda) = f_0(\lambda) 
    + \sum_{\ell=1}^N E_\ell f_\ell(\lambda), 
\eeq
and the separation variables comprise 
$N$ pairs $(\lambda_1,z_1),\ldots,(\lambda_N,z_N)$ 
of variables that satisfy the equations 
\beq
  z_j = A(\lambda_j), \quad j = 1,\ldots,N. 
\eeq
These equations, in turn, determines 
the first integrals $H_\ell$ as a function 
$E_\ell = H_\ell(\lambda_1,\ldots,\lambda_N,z_1,\ldots,z_N)$ 
of the separation variables.  More precisely, 
we assume that the non-degeneracy condition 
\beq
  \Delta = \det(f_k(\lambda_j)\mid j,k=1,\ldots,N) \not= 0 
\eeq
is satisfied.  One can then use Cramer's formula 
to solve the foregoing equations for the coefficients 
$E_l$ as 
\beq
  E_\ell = \Delta_\ell/\Delta, 
  \label{eq:Eell-Cramer}
\eeq
where  $\Delta_\ell$ denotes the determinant 
in which the $\ell$-th row of $\Delta$ is replaced 
as $a_\ell(\lambda_j) \to z_j - a_0(\lambda_j)$.  
This gives an explicit formula of the first integrals.  

For illustration, we consider the case of the open 
Toda molecule in detail below.

\section{Example: open Toda molecule}

\subsection{$L$-matrix and spectral curve}

The open Toda molecule of length $N$ 
has the Hamiltonian 
\beq
  H = \frac{1}{2}\sum_{j=1}^N p_j^2 
      + \sum_{j=2}^N e^{q_{j-1}-q_j}. 
\eeq
A Lax representation with a spectral parameter $z$ 
is presented by Krichever and Vaninsky \cite{bib:Kr-Va}.  
The $L$-matrix reads 
\beq
  L(z) = \sum_{j=2}^N c_jE_{j,j-1} 
    + \sum_{j=1}^N b_jE_{j,j} 
    + \sum_{j=2}^N E_{j-1,j} + zE_{N,1}, 
\eeq
where $E_{j,k}$ stands for the matrix unit 
with the $(j,k)$ element equal to $1$ and 
the others equal to $0$; $b_j$ and $c_j$ 
are the Flaschka variables,  
$b_j = p_j$ and $c_j = c^{q_{j-1}-q_j}$.  
Note that $L = L(0)$ is the usual tri-diagonal 
$L$-matrix without spectral parameter.   

This $L$-matrix can be derived from the $L$-matrix 
$L_{\mathrm{cl}}(z)$ of the closed Toda molecule 
(namely, the Toda lattice with periodic boundary 
condition $q_{j+N} = q_j$, $p_{j+N} = p_j$) as follows.  
The Hamiltonian of the closed Toda molecule is the sum 
\beq
  H_{\mathrm{cl}} = H + c_1, \quad 
  c_1 = e^{q_N - q_1} 
\eeq
of the Hamiltonian of the open Toda molecule 
and the potential between $q_N$ and $q_1$.  
The associated $L$-matrix is also the sum of two pieces: 
\beq
  L_{\mathrm{cl}}(z) = L(z) + c_1z^{-1}E_{1,N}. 
\eeq
In the limit as the coupling between $q_N$ and 
$q_1$ is turned off ($c_1 \to 0$), this $L$-matrix 
becomes the aforementioned matrix $L(z)$.  

This correspondence carries over to the spectral curve 
as well.  The characteristic polynomial of 
$L_{\mathrm{cl}}(z)$ takes the well known form
\beq 
  \det(\lambda I - L_{\mathrm{cl}}(z)) 
  = z + c_1 \cdots c_N z^{-1} - P(\lambda), 
\eeq
where $P(\lambda)$ is a polynomial of degree $N$.  
The spectral curve defined by this polynomial is 
a hyperelliptic curve of genus $N - 1$.  In the limit 
as $c_1 \to 0$, the characteristic polynomial reduces to 
\beq
  \det(\lambda I - L(z)) = z - A(\lambda), 
\eeq
where $A(\lambda)$ is nothing but 
the characteristic polynomial of $L = L(0)$: 
\beq
  A(\lambda) = \det(\lambda I - L). 
\eeq
The spectral curve of the open Toda molecule 
thus turns out to be the graph of $A(\lambda)$.

\subsection{Separation variables}

Having the $L$-matrix $L(z)$, one can obtain 
separation variables from poles of a suitably 
normalized eigenvector of $L(z)$.  This is 
indeed achieved in the work of Krichever and 
Vaninsky \cite{bib:Kr-Va}.  One can derive 
the same set of separation variables from 
those of the periodic Toda molecule; 
all necessary data for the latter can be 
found in the work of Flaschka and 
McLaughlin \cite{bib:Fl-Mc}.  
We here present yet another approach based on 
Moser's classical method for solving the open 
Toda molecule \cite{bib:Mo-TM}. 

Moser's method uses one of the diagonal elements 
of the resolvent $(\lambda I - L)^{-1}$: 
\beq
  R(\lambda) = \tp{\ee_N}(\lambda I - L)^{-1}\ee_N, \quad 
  \tp{\ee_N} = (0,\ldots,0,1). 
\eeq
This is a rational function of the form 
\beq
  R(\lambda) = B(\lambda)/A(\lambda).   
\eeq
Since the numerator $B(\lambda)$ is a polynomial 
of degree $N-1$, $R(\lambda)$ can be expanded 
to partial fractions as 
\beq
  R(\lambda) = \sum_{j=1}^N \frac{\rho_j}{\lambda - \alpha_j}, 
  \quad 
  A(\lambda) = \prod_{j=1}^N(\lambda - \alpha_j).  
\eeq
Moser discovered that the new dynamical variables 
$\alpha_j$ and $\rho_j$ obey the simple evolution 
equations 
\beq
  \dot{\alpha}_j = 0, \quad 
  \dot{\rho}_j = \alpha_j\rho_j, 
\eeq
which can be readily solved as $\alpha_j(t) = \alpha_j(0)$, 
$\rho_j(t) = e^{\alpha_j(0)t}\rho_j$.  The $L$-matrix 
can be reproduced from $\alpha_j$ and $\rho_j$ by 
the Stieltjes method of continued fraction.  
This is the way how Moser solved the open Toda molecule.  

From our point of view, the roots of $B(\lambda)$ 
are a half of separation variables: 
\beq
  B(\lambda) = \prod_{k=1}^{N-1}(\lambda - \lambda_k). 
\eeq
The other half $z_1,\ldots,z_{N-1}$ are given by 
the values $z_k = A(\lambda_k)$ of $A(\lambda)$ 
at the roots of $B(\lambda)$.  The Poisson brackets 
of these variables turn out to take the form 
\beq
  \{\lambda_j,\lambda_k\} = \{z_j,z_k\} = 0, \quad 
  \{\lambda_j,z_k\} = \delta_{jk}z_k. 
\eeq
One can derive these Poisson brackets in several 
different ways.  The most direct way is to use 
the result of Faybusovich and Gekhtman \cite{bib:Fa-Ge} 
on the Poisson brackets of $A(\lambda)$ and $B(\lambda)$.  

A comment on the number of variables will be in order. 
These $2N - 2$ separation variables $\lambda_k,z_k$ 
amount to $N-1$ of the $N$ degrees of freedom of 
the open Toda molecule.  The rest is the degree of freedom 
carried by the center of mass $Q = \sum_{j=1}^N q_j/N$ 
and the total momentum $P = \sum_{j=1}^N p_j$, 
which disappear in the center-of-mass frame.

\subsection{Many faces of symplectic form}  

We now have at least three different sets of 
dynamical variables besides the canonical ones 
$q_j,p_j$ --- Moser's variables $\alpha_j,\rho_j$, 
the separation variables $\lambda_k,z_k$, 
and the action-angle variables $E_\ell,\phi_\ell$. 
Accordingly, the symplectic form $\Omega$ of 
the system has many different expressions.  
We present their explicit form below 
(in the center-of-mass frame, for simplicity).  

Let us start from the expression 
\beq
  \Omega = \sum_{k=1}^{N-1} 
    d\log A(\lambda_k) \wedge d\lambda_k 
\eeq
in the separation variables $\lambda_k$ and 
$z_k = B(\lambda_k)$.  Substituting 
$A(\lambda_k) = \prod_{j=1}^N (\lambda_k - \alpha_j)$ 
leads to an intermediate expression 
\beq
  \Omega = \sum_{k=1}^{N-1} \sum_{j=1}^N 
    \frac{d\lambda_k \wedge d\alpha_j}{\lambda_k - \alpha_j}, 
\eeq
which can be reorganized to the expression 
\beq
  \Omega = \sum_{j=1}^N 
    d\log B(\alpha_j) \wedge d\alpha_j 
\eeq
in the new variables $\alpha_j$ and $B(\alpha_j)$.  
As Faybusovich and Gekhtman point out \cite{bib:Fa-Ge}
(see also Vaninsky's recent paper\cite{bib:Va-AH}), 
this is exactly the symplectic form that Atiyah and 
Hitchin \cite{bib:At-Hi} introduced on the moduli 
space of $\SU(2)$ monopoles.  In terms of Moser's variables 
$\alpha_j$ and $\rho_j = B(\alpha_j)/A'(\alpha_j)$, 
the last expression reads 
\beq
  \Omega = \sum_{j=1}^N d\log\rho_j \wedge d\alpha_j 
    + \sum_{j\not= k}\frac{d\alpha_j 
        \wedge d\alpha_k}{\alpha_j - \alpha_k}. 
\eeq

An expression in action-angle variables can be 
obtained as follows.  Let $E_\ell$ be the 
coefficients of $A(\lambda)$: 
\beq
  A(\lambda) = \lambda^N 
    + \sum_{\ell=2}^N E_\ell\lambda^{N-\ell}. 
\eeq
($E_1$ is equal to the total momentum, hence 
absent in the center-of-mass frame.)  
The foregoing expression of $\Omega$ can 
now be rewritten as 
\beq
  \Omega = \sum_{\ell=2}^N dE_\ell \wedge d\phi_\ell, 
\eeq
where 
\beq
  \phi_\ell = \sum_{k=1}^{N-1} \int^{\lambda_k} 
    \frac{\lambda^{N-\ell}}{A(\lambda)}d\lambda. 
\eeq
This shows that $\phi_\ell$ is an angle variable 
conjugate to the action variable $E_\ell$.  
It is easy to see that generating function $S$ 
of the action-angle variables, too, has a similar 
simple expression: 
\beq
  S = \sum_{k=1}^{N-1} \int^{\lambda_k} 
        \log A(\lambda) d\lambda. 
\eeq

\section{Other examples}

\subsection{Open RT molecule}

The open Ruijsenaars-Toda (RT) molecule may be 
thought of as a limit of the closed RT molecule 
(namely, the RT lattice with periodic boundary 
condition $q_{j+N} = q_j$ and $p_{j+N} = p_j$).  
The passage to the open molecule is parallel 
to the case of the Toda lattice. The closed molecule 
has the Hamiltonian 
\beq
  H_{\mathrm{cl}} = 
    \sum_{j=1}^N e^{p_j}(1 + e^{q_{j-1}-q_j})^{1/2}
      (1 + e^{q_j-q_{j+1}})^{1/2}, 
\eeq
which turns into the Hamiltonian $H$ of the 
open molecule as the potential $e^{q_N - q_1}$ 
between $q_1$ and $q_N$ is turned off.  

SOV of this case is more delicate than the Toda 
molecules, because curves arising here are NOT 
a spectral curve in the usual sense.  This is 
already the case for the closed molecule. 
The closed RT molecule was solved by Bruschi 
and Ragnisco \cite{bib:Ra-Br} by an 
algebro-geometric method, but they had to 
to use an unusual Lax representation with 
a matrix $L(\lambda,z)$ that depends on 
two parameters $\lambda$ and $z$.  Since 
$L(\lambda,z)$ amounts to $L(z) - \lambda I$ in 
the usual Lax representation, they considered 
the curve defined by the equation 
\beq
  f(\lambda,z) = \det L(\lambda,z) = 0, 
\eeq
and obtained a set of separation variables as 
Flaschka and McLaughlin \cite{bib:Fl-Mc} did for 
the closed Toda molecule.  Therefore, bearing in mind 
the case of the Toda molecules, it is not difficult 
to see what appears in the limit to the open RT molecule. 

In the limit to the open molecule, as expected, 
the hyperelliptic curve of Bruschi and Ragnisco 
degenerates to the graph of a polynomial $A(\lambda)$ 
of the form 
\beq
  A(\lambda) = \lambda^N 
    + \sum_{\ell=1}^{N-1} E_\ell \lambda^{N-\ell} + 1. 
\eeq
Separation variables $\lambda_1,\ldots,\lambda_{N-1}$ 
are the roots of a polynomial $B(\lambda)$ of degree $N-1$, 
and their conjugate variables $z_1,\ldots,z_{N-1}$ are 
defined by $z_k = A(\lambda_k)$.  The Poisson brackets 
are slightly different from those of the Toda molecules: 
\beq
  \{\lambda_j,\lambda_k\} = \{z_j,z_k\} = 0, \quad 
  \{\lambda_j,z_k\} = \delta_{jk}\lambda_k z_k. 
\eeq
The symplectic form $\Omega$ of the system, too, 
is modified as 
\beq
  \Omega 
  &=& \sum_{k=1}^{N-1}d\log A(\lambda_k)\wedge d\log\lambda_k 
  \nonumber \\
  &=& \sum_{j=1}^N d\log B(\lambda_j)\wedge d\log\alpha_j 
  \nonumber \\
  &=& \sum_{\ell=1}^{N-1} dE_\ell \wedge d\phi_\ell, 
\eeq
where $\alpha_j$'s are the roots of $A(\lambda)$, 
and $\phi_\ell$'s are given by 
\beq
  \phi_\ell = 
    \sum_{k=1}^{N-1} \int^{\lambda_k} 
    \frac{\lambda^{N-\ell}}{A(\lambda)}d\log\lambda. 
\eeq

\subsection{Rational and trigonometric CM systems}

The rational and trigonometric Calogero-Moser (CM) 
systems may be thought of as degeneration of 
the elliptic CM system.  This allows us to find 
a Lax representation with a spectral parameter 
for these systems. 

Let us recall the following $L$-matrix \cite{bib:Kr} 
for the elliptic CM system: 
\beq
  L(z) = \sum_{j=1}^N p_jE_{j,j} 
    + \sum_{j\not= k}gi\phi(q_j - q_k, z)E_{j,k}, 
\eeq
where $\phi(u,z)$ is written in terms of 
the Weierstrass $\sigma$ function as 
\beq
  \phi(u,z) = \frac{\sigma(u + z)}{\sigma(u)\sigma(z)}. 
\eeq
As the elliptic curve degenerates to a rational curve, 
$\phi(u,z)$ turns into a trigonometric (hyperbolic) 
function 
\beqnn
  \phi(u,z) \to 
  \frac{\sinh(u + z)}{\sinh(u)\sinh(z)} = \coth(u) + \coth(z)
\eeqnn
or a rational function 
\beqnn
  \phi(u,z) \to 
  \frac{u + z}{uz} = \frac{1}{u} + \frac{1}{z}. 
\eeqnn
Remakably, the limit of $\phi(u,z)$ splits 
into two functions of a single variable.  
Accordingly, the limit of $L(z)$ becomes 
the sum of two matrices as 
\beq
  L(z) \to 
  \left\{\begin{array}{ll}
  L + gi\coth(z)K & \mbox{(trigonometric)}\\
  L + giz^{-1}K & \mbox{(rational)}
  \end{array}\right. 
  \label{eq:L(z)-limit}
\eeq
where 
\beq
  L = \sum_{j=1}^N p_jE_{j,j} 
    + \sum_{j\not= k}gi\coth(q_j - q_k)E_{j,k} 
\eeq
in the trigonometric limit, 
\beq
  L = \sum_{j=1}^N p_jE_{j,j} 
    + \sum_{j\not= k}gi(q_j - q_k)^{-1}E_{j,k} 
\eeq
in the rational limit, and 
\beq
  K = \sum_{j\not= k}E_{j,k} 
\eeq
for both cases.  $L$ is nothing but the usual 
$L$-matrix of the trigonometric and rational 
CM systems; $K$ is the constant matrix that 
appears in the method of Hamiltonian reduction 
(or the ``projection method'') for solving 
these systems \cite{bib:Ol-Pe,bib:Ka-Ko-St}. 
The linear combination of $L$ and $K$ in 
the limit (\ref{eq:L(z)-limit}) of $L(z)$ 
gives an $L$-matrix with a spectral parameter 
for the rational and trigonometric CM systems 

We modify these rational and trigonometric $L$-matrices 
slightly.  Note that $K$ can be expressed as 
\beq
  K = \ee\tp{\ee} - I, \quad 
  \tp{\ee} = (1,\ldots,1). 
\eeq
Therefore the term proportional to $K$ in 
(\ref{eq:L(z)-limit}) is a linear combination of 
a rank-one matrix and a scalar matrix; the latter 
does not affect to the Lax equation.  We can thus 
drop the scalar term and {\it redefine} the $L$-matrix 
for the degenerate CM systems as 
\beq
  L(z) = L + gi\coth(z)\ee\tp{\ee}
\eeq
for the trigonometric case and 
\beq
  L(z) = L + giz^{-1}\ee\tp{\ee} 
\eeq
for the rational case.  

We now apply the so called Weinstein-Aronszajn formula 
\beqnn
  \det(M + \xx\tp{\yy}) = (1 + \tp{\yy}M^{-1}\xx) \det M 
\eeqnn
to the characteristic polynomial of $L(z)$. 
This yields the expression 
\beq
  \det(\lambda I - L(z)) = 
  \left\{\begin{array}{ll} 
  P_0(\lambda) - gi\coth(z)P_1(\lambda) & \mbox{(trigonometric)} \\
  P_1(\lambda) - giz^{-1}P_1(\lambda) & \mbox{(rational)} 
  \end{array}\right.
\eeq
where 
\beq
  P_0(\lambda) = \det(\lambda I - L), \quad 
  P_1(\lambda) = \tp{\ee}\widetilde{(\lambda I - L)}\ee. 
\eeq
($\widetilde{M}$ denote the cofactor matrix of $M$.) 
The equation of the spectral curve of $L(z)$ 
thus turns out to take the form 
\beq
  z = \frac{P_1(\lambda)}{P_0(\lambda)} 
\eeq
for the rational case and 
\beq
  e^{2z} = 
    \frac{P_0(\lambda) + P_1(\lambda)}
         {P_0(\lambda) - P_1(\lambda)} 
\eeq
for the trigonometric form.  This is what we have sought. 

It should be mentioned that the same equation of 
the spectral curve for the trigonometric case was 
also derived by Vaninsky \cite{bib:Va-CM} from a different 
point of view.  We can proceed further to the construction 
of separation coordinates, though we omit details here.

\subsection{Rational and trigonometric RS systems}

The rational and trigonometric Ruijsenaars-Schneider 
(RS) systems can be treated in the same way as 
the CM systems.  We again start from the elliptic 
RS system, and consider the limit as the elliptic curve 
degenerates to a rational curve.  It is convenient 
to use the following $L$-matrix of the elliptic RS system 
(which is gauge equivalent to the $L$-matrix of Bruschi 
and Calogero \cite{bib:Br-Ca}): 
\beq
  L(z) = \sum_{j,k=1}^N (h_j h_k)^{1/2} 
         \phi(q_j - q_k + \gamma, z) E_{j,k}, 
\eeq
where $\gamma$ is a constant and $h_j$'s are fairly 
complicated functions of the canonical variables.  
In the limit to the trigonometric or rational models,  
the $L$-matrix again splits into the sum of two matrices, 
one being the $L$-matrix $L$ without spectral parameter, 
and the other a rank-one matrix.  One can thus use 
the Weinstein-Aronszajn formula again to show that 
the spectral curve is the graph of a function $A(\lambda)$.  
The same curve was derived by Braden and Marshakov 
\cite{bib:Br-Ma} in the context of supersymmetric 
gauge theories.

\section{Variants and generalizations}

The goal of this research is to search for 
new integrable systems from the point of view 
of SOV.  Let us show a few results below.  

\subsection{Trigonometric and elliptic analogues}

Apart from the CM and RS systems, the foregoing 
examples are associated with a pair of polynomials 
$A(\lambda)$ and $B(\lambda)$.  The main result of 
the joint work with Takebe \cite{bib:Ta-Ta} 
is to construct the following analogues defined 
on a cylinder and a torus: 
\begin{enumerate}
\item Trigonometric version 
\beq
  A(\lambda) = \prod_{j=1}^N \sinh(\lambda - \alpha_j), \quad 
  B(\lambda) = \prod_{j=1}^N \sinh(\lambda - \lambda_j). 
\eeq
\item Elliptic version 
\beq
  A(\lambda) = \prod_{j=1}^N \sigma(\lambda - \alpha_j), \quad 
  B(\lambda) = \prod_{j=1}^N \sigma(\lambda - \lambda_j). 
\eeq
\end{enumerate}
The construction can be further extended to the case 
where $A(\lambda)$ and $B(\lambda)$ have a different 
number of zeroes.

\subsection{Elliptic fibration}

Another possible elliptic analogue is to modify 
the $z$ direction of the $(\lambda,z)$ space to 
an elliptic curve.   A naive way will be to replace 
\beq
  \Omega = \sum_{j=1}^N d\log z_j \wedge d\lambda_j \ \to \ 
  \Omega = \sum_{j=1}^N \frac{dz_j}{y_j} \wedge d\lambda_j 
\eeq
where $y_j$ and $z_j$ are constrained 
by the equation 
\beqnn
  y_j^2 = 4z_j^3 + g_2 z_j + g_3 
\eeqnn
of an elliptic curve.  Actually, one can go 
further by allowing the coefficients  
$g_2$ and $g_3$ to depend on $\lambda_j$ as 
\beq
  y_j^2 = 4z_j^3 + g_2(\lambda_j) z_j + g_3(\lambda_j), 
\eeq
where $g_2(\lambda)$ and $g_3(\lambda)$ are 
functions (e.g., polynomials) of $\lambda$.  
Geometrically, this is the structure called 
an ``elliptic fibration''.  Integrable systems 
of this type, too, are considered in the joint work 
with Takebe \cite{bib:Ta-Ta}.  

\subsection{Higher genera} 
Generalization to a complex algebraic curve $C_0$ 
of genus $g > 1$ (and a Riemann surface of more 
general type) will be achieved in several 
different ways.  Let us propose an idea below.  

The idea is to choose the $(A,B)$ pair as 
nonzero elements of the vector space 
$\calL(R_1 + \cdots + R_N)$ of meromorphic 
functions $f(P)$ ($P \in C_0$) on $C_0$ 
with at most simple poles at $N$ fixed points 
$R_1,\ldots,R_N$.  If $N > 2g - 2$ (or if $N > g$ 
and $R_1,\ldots,R_N$ are in general position), 
the Riemann-Roch theorem shows that this vector space 
is spanned by $N - g + 1$ linearly independent elements 
$f_0(P),f_1(P),\ldots,f_{N-g}(P)$.  We write 
$A(P)$ and $B(P)$ by this basis as 
\beq
  A(P) = \sum_{\ell=0}^{N-g} u_\ell f_\ell(P), \quad 
  B(P) = \sum_{\ell=0}^{N-g} v_\ell f_\ell(P) 
\eeq
and view $(u_\ell)_{\ell=0}^{N-g}$ and 
$(v_\ell)_{\ell=0}^{N-g}$ as homogeneous 
coordinates on two copies of the projective space 
$\PP^{N-g}$.  The $(A,B)$ pair is thus parametrized 
by $2N - 2g$ ``moduli'', which play the role of 
dynamical variables.  
Now let us choose a holomorphic differential 
$d\lambda$ on $C_0$, and consider its (multi-valued) 
primitive function 
\beq
  \lambda(P) = \int^P d\lambda. 
\eeq
In this quite general setup, one can use a trick 
based on the residue theorem \cite{bib:Ta-Ta} to 
derive the identity 
\beq
    \sum_{j=1}^N d\log B(P_j)\wedge d\lambda(P_j) 
  = \sum_{j=1}^N d\log A(Q_j)\wedge d\lambda(Q_j), 
\eeq
where $P_j$ and $Q_j$ are zeroes of $A(P)$ and 
$B(P)$, respectively; $d$ is understood to be 
the total differential in the $2N - 2g$ moduli 
of the $(A,B)$ pair.  This defines a symplectic form 
$\Omega$ on the moduli space.  The inhomogeneous 
coordinates $H_\ell = u_\ell/u_0$ give a set of 
functions in involution with respect to $\Omega$, 
thereby defining an integrable system.  The role of 
separation variables is played by 
\beq
  z_j = A(Q_j), \quad \lambda_j = \lambda(Q_j), 
\eeq
though they are not independent variables.

\subsection*{Acknowledgements}
I would like to thank Takashi Takebe for collaboration 
for years, Evgueni Sklyanin for valuable comments 
in an early stage of this research, and Dmitry Korotkin 
and Jean-Pierre Fran\c{c}oise for discussion and comments 
at the workshop.  This work was partly supported by 
the Grant-in-Aid for Scientific Research 
(No. 12640169 and No. 14540172) from the Ministry of 
Education, Culture, Sports and Technology.

\end{document}